\documentclass[twocolumn]{aastex62}

\usepackage[utf8]{inputenc}
\usepackage{xspace}
\usepackage{url}
\usepackage{listings}
\usepackage[T1]{fontenc} 
\usepackage{textcomp} 
\usepackage{inconsolata} 
\newcommand{\package}[1]{\texttt{#1}\xspace}
\newcommand{\astroquery}{\package{astroquery}}
\newcommand{\astropy}{Astropy\xspace}
\newcommand{\astropypkg}{\package{astropy}}

\definecolor{dkgreen}{rgb}{0,0.6,0}
\definecolor{gray}{rgb}{0.5,0.5,0.5}
\definecolor{mauve}{rgb}{0.58,0,0.82}

\begin{document}
\newcommand{\nraojansky}{\affiliation{\it{Jansky fellow of the National Radio Astronomy Observatory, 1003 Lopezville Rd, Socorro, NM 87801 USA }}}
\newcommand{\stsci}{\affiliation{Space Telescope Science Institute, 3700 San Martin Dr, Baltimore, MD 21218, USA}}
\newcommand{\gsfc}{\affiliation{NASA Goddard Space Flight Center, Astrochemistry Laboratory, 8800 Greenbelt Road, Greenbelt, MD 20771, USA}}
\newcommand{\gsoc}{\affiliation{Google Summer of Code Student}}

\author[0000-0001-6431-9633]{Adam Ginsburg}
\nraojansky

\correspondingauthor{Adam Ginsburg}
\email{aginsbur@nrao.edu; adam.g.ginsburg@gmail.com}

\author[0000-0002-3713-6337]{Brigitta M. Sip\H{o}cz}
\affiliation{DIRAC Institute, Department of Astronomy, University of Washington, 3910 15th Avenue NE, Seattle, WA 98195, USA}
\affiliation{Institute of Astronomy, University of Cambridge, Madingley Road, Cambridge, CB3 0HA, UK}

\author[0000-0002-9314-960X]{C. E. Brasseur}
\stsci

\author[0000-0002-2478-6939]{Philip S. Cowperthwaite}
\affiliation{Harvard-Smithsonian Center for Astrophysics, 60 Garden St., Cambridge, MA 02138, USA}

\author[0000-0001-7988-8919]{Matthew W. Craig}
\affiliation{Department of Physics and Astronomy, Minnesota State University Moorhead, 1104 7th Ave S., Moorhead, MN 56563, USA}

\author[0000-0002-4198-4005]{Christoph Deil}
\affiliation{Max-Planck-Institut f\"{u}r Kernphysik, Heidelberg, Germany}

\author[0000-0002-9809-8215]{James Guillochon}
\affiliation{Harvard-Smithsonian Center for Astrophysics, 60 Garden St., Cambridge, MA 02138, USA}

\author[0000-0001-6340-8220]{Giannina Guzman}
\affiliation{Department of Astrophysics and Planetary Science, Villanova University, 800 East Lancaster Avenue, Villanova, PA 19085, USA}
\gsfc

\author{Simon Liedtke}
\gsoc

\author[0000-0003-0079-4114]{Pey Lian Lim}
\stsci

\author[0000-0002-8130-1440]{Kelly E. Lockhart}
\affiliation{Harvard-Smithsonian Center for Astrophysics, 60 Garden St., Cambridge, MA 02138, USA}

\author[0000-0002-8132-778X]{Michael Mommert}
\affiliation{Lowell Observatory, 1400 W Mars Hill Rd, Flagstaff, AZ 86001, USA}

\author[0000-0003-2528-3409]{Brett M. Morris}
\affiliation{Astronomy Department, University of Washington, Seattle, WA 98195, USA}

\author[0000-0002-8898-4015]{Henrik Norman}
\affiliation{Winter Way, Uppsala, Sweden}
\affiliation{ESAC Science Data Centre, European Space Agency, Madrid, Spain}

\author{Madhura Parikh}
\gsoc

\author[0000-0002-1100-5734]{Magnus V. Persson}
\affiliation{Department of Space, Earth and Environment, Chalmers University of Technology, Onsala Space Observatory, 439 92, Onsala, Sweden}

\author[0000-0002-8642-1329]{Thomas P. Robitaille}
\affiliation{Aperio Software Ltd., Headingley Enterprise and Arts Centre, Bennett Road, Leeds, LS6 3HN, United Kingdom}

\author{Juan-Carlos Segovia}
\affiliation{ESAC Science Data Centre, European Space Agency, Madrid, Spain}

\author[0000-0001-9898-5597]{Leo P. Singer}
\affiliation{Astroparticle Physics Laboratory, NASA Goddard Space Flight Center, Mail Code 661, Greenbelt, MD 20771, USA}
\affiliation{Joint Space-Science Institute, University of Maryland, College Park, MD 20742, USA}

\author[0000-0002-9599-310X]{Erik J. Tollerud}
\stsci

\author[0000-0002-0455-9384]{Miguel de Val-Borro}
\gsfc
\affiliation{Department of Physics, Catholic University of America, Washington, DC 20064, USA}

\author[0000-0001-9930-7886]{Ivan Valtchanov}
\affiliation{European Space Astronomy Centre, European Space Agency, Madrid, Spain}

\author[0000-0002-2958-4738]{Julien Woillez}
\affiliation{European Southern Observatory, Karl-Schwarzschild-Str. 2, 85748 Garching bei M\"{u}nchen, Germany}

\author{the Astroquery collaboration, a subset of the astropy collaboration}

\title{\astroquery: An Astronomical Web-Querying Package in Python}

\begin{abstract}
\astroquery is a collection of tools for requesting data from databases hosted
on remote servers with interfaces exposed on the internet, including those with
web pages but without formal application program interfaces (APIs).  These
tools are built on the Python
\package{requests} package,
which is used to make HTTP requests, and
\astropypkg, which provides most of the data parsing functionality.
\astroquery modules generally attempt to replicate the web page interface
provided by a given service as closely as possible, making the transition
from browser-based to command-line interaction easy.
\astroquery
has received significant contributions from throughout the astronomical community,
including several significant contributions from telescope archives.
\astroquery enables the creation of fully reproducible workflows from data
acquisition through publication.  This paper describes the philosophy, basic
structure, and development model of the \astroquery package.
The complete documentation for astroquery can be found at
\url{http://astroquery.readthedocs.io/}.
\footnote{%
The repository associated with this paper is:\\
\url{https://github.com/adamginsburg/astroquery-paper}
}
\end{abstract}

\section{Introduction}
In the past few decades, large-scale surveys have played a huge role in
advancing our understanding of the universe, and these surveys have produced
enormous reservoirs of data that astronomers regularly access.  However, tools
for accessing these reservoirs are heterogeneous and often only available via
graphical user interfaces (GUIs) or web sites.

One of the cornerstones of research is reproducibility. To be able to reproduce
research, the data need to be available to everyone. Many scientific journals
encourage or demand that the underlying data  accompany the article or be
uploaded to a hosting service. Data sharing is not only important for new
results, but also to provide the ability to test and verify published results.
While many different efforts to promote data sharing have made the practice
more common, it is difficult to keep track of how and where to retrieve a given
data set. A common scripted interface to tie all these services together is a
good way to make all the different data more accessible, and it provides
authors with the ability to make the full analysis process they used -- from
data download to publication -- repeatable.  A centrally maintained library
also safeguards against inevitable `link rot' on data archives, moving some of
the responsibility for maintaining long-term reproducibility from each
individual researcher to the broader community.

Data sharing has taken on a variety of forms.  The most prominent are the
major observatory archives: MAST, NOAO, ESO, ESA, IPAC, CDS, NRAO, CXC, HEASARC,
and CADC are the main organizations hosting raw and processed data from
ground and space based telescopes.  These data archives also serve as the
primary means for serving data to users when the data are taken in queue
mode, i.e., when the data are taken while the observer is not on-site.

\begin{deluxetable*}{lp{8.5cm}ll}
  \tablecaption{List of all Services \& Surveys \astroquery modules support.}
  \label{tab:surveys}
  \tablehead{Module name & Service or Organization & URL}
  \startdata
  \package{alfalfa} &  ALFALFA data repository & \url{http://arecibo.tc.cornell.edu/hiarchive/alfalfa} \\
  \package{alma} &  Atacama Large Millimeter/submillimeter Array Archive & \url{http://almascience.org} \\
  \package{atomic} &  Atomic Line List & \url{http://www.pa.uky.edu/~peter/atomic} \\
  \package{besancon} & Besancon model of the Galaxy& \url{http://model.obs-besancon.fr} \\
  \package{cds}  & Centre de Données astronomiques de Strasbourg & \url{http://cds.u-strasbg.fr} \\
  \package{cosmosim} & CosmoSim database & \url{https://www.cosmosim.org/uws/query} \\
  \package{esasky} &  ESASky of the European Space Agency & \url{http://sky.esa.int} \\
  \package{eso} & European Southern Observatory Science Archive & \url{http://archive.eso.org/cms.html} \\
  \package{exoplanet\_orbit\_database} & Exoplanet Orbit Database& \url{http://exoplanets.org}\\
  \package{fermi} & Fermi Gamma-ray Space Telescope Data & \url{https://fermi.gsfc.nasa.gov/ssc/data} \\
  \package{gaia} & Gaia Archive of the European Space Agency & \url{https://gea.esac.esa.int/archive} \\
  \package{gama} & Galaxy and Mass Assembly Survey & \url{http://www.gama-survey.org/dr2/query} \\
  \package{heasarc} &  High Energy Astrophysics Science Archive Research Center & \url{https://heasarc.gsfc.nasa.gov} \\
  \package{hitran} & HIgh-resolution TRANsmission molecular absorption database & \url{http://hitran.org/hapi} \\
  \package{ibe} & IRSA Image Server & \url{http://irsa.ipac.caltech.edu/ibe} \\
  \package{irsa} & IRSA Catalog Query Service & \url{https://irsa.ipac.caltech.edu} \\
  \package{irsa\_dust} & IRSA Galactic Dust Reddening and Extinction Query & \url{https://irsa.ipac.caltech.edu/applications/DUST} \\
  \package{jplhorizons} & JPL's HORIZONS system  & \url{https://ssd.jpl.nasa.gov/horizons_batch.cgi} \\
  \package{jplsbdb} &JPL's Small-Body DataBase  & \url{https://ssd-api.jpl.nasa.gov/doc/sbdb.html} \\
  \package{jplspec} &JPL's Spectral Catalog  & \url{https://spec.jpl.nasa.gov/cgi-bin/catform} \\
  \package{lamda} & Leiden Atomic and Molecular Database & \url{http://home.strw.leidenuniv.nl/~moldata} \\
  \package{magpis} & The Multi-Array Galactic Plane Imaging Survey & \url{https://third.ucllnl.org/gps} \\
  \package{mast} & Barbara A. Mikulski Archive for Space Telescopes & \url{https://mast.stsci.edu} \\
  \package{mpc} & Minor Planet Center Ephemeris Service & \url{https://minorplanetcenter.net} \\
  \package{nasa\_ads} & SAO/NASA Astrophysics Data System & \url{https://api.adsabs.harvard.edu} \\
  \package{nasa\_exoplanet\_archive} &NASA Exoplanet Archive  & \url{https://exoplanetarchive.ipac.caltech.edu} \\
  \package{ned} & NASA Extragalactic Database & \url{https://ned.ipac.caltech.edu} \\
  \package{nist} &NIST Atomic Spectra Database  & \url{https://physics.nist.gov/PhysRefData/ASD} \\
  \package{nrao} &  National Radio Astronomy Observatory Data Archive & \url{https://archive.nrao.edu/archive}\\
  \package{nvas} & NRAO VLA Archive Survey Images Page & \url{https://archive.nrao.edu/nvas} \\
  \package{oac} & Open Astronomy Catalog & \url{https://astrocats.space} \\
  \package{ogle} & Interstellar Extinction toward the Galactic Bulge from OGLE-III data & \url{http://ogle.astrouw.edu.pl/cgi-ogle/getext.py} \\
  \package{open\_exoplanet\_catalogue} & Open Exoplanet Catalogue & \url {http://openexoplanetcatalogue.com} \\
  \package{sdss} & Sloan Digital Sky Survey & \url{http://skyserver.sdss.org}\\
  \package{sha} & Spitzer Heritage Archive & \url{http://sha.ipac.caltech.edu/applications/Spitzer/SHA } \\
  \package{simbad} & CDS SIMBAD Astronomical Database & \url{http://simbad.u-strasbg.fr}\\
  \package{skyview} & NASA's SkyView Query  & \url{ http://skyview.gsfc.nasa.gov} \\
  \package{splatalogue} & Splatalogue Database for astronomical spectroscopy query & \url{https://www.cv.nrao.edu/php/splat} \\
  \package{ukidss} & UKIRT Infrared Deep Sky Survey & \url{http://wsa.roe.ac.uk}\\
  \package{vamdc} & VAMDC molecular line database & \url{https://vamdclib.readthedocs.io/} \\
  \package{vizier} & CDS VizieR Astronomical Catalogues & \url{http://vizier.u-strasbg.fr} \\
  \package{vo\_conesearch} &Simple Cone Search Databases & \url{https://astropy.stsci.edu/aux/vo_databases} \\
  \package{vsa} & Vista Science Archive & \url{http://vsa.roe.ac.uk}\\
  \package{xmatch} & CDS X-Match Service & \url{http://cdsxmatch.u-strasbg.fr} \\
  \enddata
\end{deluxetable*}

In addition to observatories and telescopes, individual surveys often share
their full data sets.  In some cases, these data sets are shared via the
observatory that acquired them, for example, the all-sky data acquired with
Planck, WMAP, and COBE\@.  Other surveys, particularly ground-based surveys,
serve their own data.  Examples include SDSS, 2MASS, UKIDSS, and VSA.


Individual teams and small groups often share their data via their own
custom websites.  These services do not follow any particular standard and can
be widely varied in the type and amount of data shared.  Sometimes these data
are shared via the archive systems (e.g., IRSA at IPAC hosts many individual
survey data sets), while others use their own web hosting systems (e.g.,
MAGPIS).

Finally, there are other data types relevant to astronomy that are not
served by the typical astronomical databases.  Examples include databases of
molecular and atomic properties, such as those provided by Splatalogue and
the NIST Atomic Spectra Database, bibliographic databases such as
the NASA Astrophysics Data System (ADS), or services that are computationally
intensive or require regular updates, like Solar System ephemerides
provided by services like JPL HORIZONS, or the Minor Planet Center.

\astroquery arose from a desire to access these databases from the Python
command line in a scriptable fashion.  Script-based data access provides
astronomers with the ability to make reproducible analysis scripts and
pipelines in which the data are retrieved and processed into scientifically
relevant results with minimal user interaction.

In this paper, we provide an overview of the \astroquery package.  Section
\ref{sec:software} describes the basic layout of the software and the shared
API concept underlying all modules.  Section \ref{sec:development} describes
the development model.  Finally, Section \ref{sec:documentation} describes how
\astroquery is documented.


\section{The Software}
\label{sec:software}
\astroquery consists of a collection of modules that mostly share a similar
interface, but are meant to be used independently.  They are primarily based on
a common framework that uses the Python
\package{requests}\footnote{\url{http://docs.python-requests.org/}} package to
perform HTTP requests to communicate with web services.

For new module development, there is a \texttt{template\_module} consisting of a
folder with several individual python code files that lays
out the basic framework of any new module.  All modules have a single core
\texttt{class} that has some number of \texttt{query\_*} methods.
The most common query method is \texttt{query\_region}, which usually
provides a ``cone search'' functionality, i.e., they search for data within a
circular region. The results of
the queries then are returned in an \astropypkg
\citep{Astropy-Collaboration2018, Astropy-Collaboration2013}
\texttt{Table}.\footnote{\url{http://docs.astropy.org/en/stable/table/}}

An example using the SIMBAD interface is shown
below:\footnote{\url{http://astroquery.readthedocs.io/en/latest/simbad/simbad.html}}
\begin{lstlisting}[caption=Query SIMBAD for a region around M81]
from astroquery.simbad import Simbad
result_table = Simbad.query_region("m81")
\end{lstlisting}
In this example, \texttt{Simbad} is an instance of \\
\texttt{astroquery.simbad.SimbadClass}, and \texttt{result\_table} is an
\texttt{astropy.table.Table} containing the objects near M81.
This common interface allows users to use different services and process
the resulting data in the same manner despite the differences in the underlying
methods and services (e.g., \texttt{SDSS.query\_region()},
\texttt{Simbad.query\_region()}, \texttt{NED.query\_region()}, etc.)

While there is a common suggested API described in the \texttt{template\_module},
individual packages are not \emph{required} to support this API because, for
some, it is not possible.  For example, the atomic and molecular databases refer
to physical data that is not related to positions on the sky and therefore
their \astroquery modules cannot include \texttt{query\_region} methods. The
same applies to Solar System object ephemerides queries. Differences in the API
are discussed in the \astroquery documentation (see Section
\ref{sec:documentation}).

\subsection{Version Numbers}
\label{sec:versionnumbers}
\astroquery uses the same format as traditional semantic versioning,
with versions indicated in the format \texttt{MAJOR.MINOR.PATCH.devCOMMIT\_ID} (for
example, \texttt{0.3.9.dev4581}).

\astroquery patches are frequently made to accommodate upstream changes, i.e.,
changes made to the remote service, and as such are not guaranteed to be
backward-compatible. Thus, starting in mid-2018, \astroquery switched from a
manual release model to a continuous deployment model.  Prior to this change,
the \texttt{MAJOR.MINOR.PATCH} versions were each created manually by one of
the maintainers, then pushed to package release services.  After this change,
each accepted pull request automatically triggered a new release via the python
package index.\footnote{\url{https://pypi.org/}}   We created a new manual release,
v0.3.9, to accompany the publication of this paper.

%

\subsection{HTTP User-Agent}
\label{sec:useragent}
\astroquery identifies itself to host services using the \texttt{HTTP
User-Agent} header data, which is automatically produced and sent to the
archives with every request.
Users do not need to be aware of this metadata being sent with their
queries, but the information can be used by data hosting services to determine
how many users are accessing their service via \astroquery and to assist in
debugging if improper queries are being submitted.

The format of the user agent string is:
\[\texttt{astroquery/\{version\} \{requests\_version\}}\] where
\texttt{\{version\}} is a version number of the form described in \S
\ref{sec:versionnumbers} and \texttt{\{requests\_version\}} is the
corresponding version of the Python \package{requests} package. For example:
\[\texttt{astroquery/0.3.9.dev4863 python-requests/2.14.2}\]

\subsection{The API}
The common API has a few features defined in the template module.
Each service is expected to provide the following interfaces, assuming they are
applicable:

\begin{itemize}
    \item \texttt{query\_region} - A method that accepts an \astropy
        \texttt{SkyCoord} object representing a point on the sky plus a
        specification of the radius around which to search.
        The returned object is an \astropy table.
    \item \texttt{query\_object} - A method that accepts the name of an
        object.  This method relies on the service to resolve the object name, i.e., it does not use a name resolver
        like \texttt{SESAME}.\footnote{\url{http://cds.u-strasbg.fr/cgi-bin/Sesame}}
        The returned object is an \astropy table.
    \item \texttt{get\_images} - For services that provide image data, this
        method accepts an \astropy \texttt{SkyCoord} object
        and a radius to search for data that cover the specified target. The
        returned object is a list of \texttt{astropy.io.fits.HDUList} objects.

\end{itemize}

We also require a low-level interface to the services so that queries
with very large results can be handled by other methods (e.g., data streaming)
if needed.
The low-level interface consists of a series of methods with the same
names, but with the additional suffix \texttt{\_async} (e.g.,
\texttt{query\_async}).  The
\texttt{query*\_async} methods return a \texttt{requests.Response} object
from the accessed website, providing developers with
the ability to access the data in a stream or access only the response
metadata (i.e., the \texttt{async} methods do not download the corresponding
data, so they may be useful for collecting metadata for very large files).  The
\texttt{get\_images\_async} method returns
\texttt{FileContainer} objects that similarly provide `lazy' access to the
data, but specifically for FITS files.  Contributors need only implement
these \texttt{\_async} methods because there is a wrapper tool that converts
\texttt{\_async} methods into their corresponding non-asynchronous versions.

Deviations from this standard API are documented in the \astroquery
documentation (see Section \ref{sec:documentation}).  Most deviations
are for services for which \texttt{query\_region} methods are not defined,
such as atomic and molecular line databases.

\subsection{Caching and login functionality}
Astroquery provides tools to handle multiple aspects of querying that are
common to all modules.  The \texttt{BaseQuery} metaclass provides tools for
caching requests and downloaded data, reducing the duration and the network
load for repeated queries.  Cached data are stored in the user's
\texttt{\textasciitilde/.astropy/cache/astroquery} directory.  The
\texttt{BaseQuery} metaclass is also responsible for setting the User-Agent (\S
\ref{sec:useragent}).  The \texttt{QueryWithLogin} metaclass provides a
framework for logging in securely to services that require user
authentication, including a credential storage mechanism.

\subsection{Error handling}
Some queries will inevitably fail.  Failures can take on different modes.  For
common and expected modes, such as searching for an object or location on the
sky and getting no results, the result is clearly communicated as a simple null
result or empty table.  For unpredictable and unexpected errors, such as server
failures, timeouts, and other related communication issues, the errors are handled
by the \texttt{requests} module, and normal HTTP responses are returned (e.g.,
HTTP 200 means the request was successful, while 503 indicates the request
was forbidden by server-side permissions;  a complete list can be found at
\url{https://en.wikipedia.org/wiki/List_of_HTTP_status_codes}).

In some cases, when we know a particular failure mode is likely (because the
developers have encountered it at least once), we catch and raise a specific
\texttt{Exception} or \texttt{Warning}.  The full list of these is in the
\texttt{exceptions.py} file.  Developers can use these custom exceptions
to build in additional robustness to data pipelines using astroquery
by either implementing workarounds to known issues or correctly informing
users of the problem.

\subsection{Testing}
Astroquery testing is somewhat different from most other packages in the
scientific Python
ecosystem.  While the tests are based on the \astropy testing infrastructure and use
pytest to run and check the outputs, the astroquery tests are split into
\emph{remote} and \emph{local}.  The remote tests exactly replicate what a user
would enter at the command line, but they are dependent on the stability of the
remote services.

In our experience it is quite rare for all of the astroquery-supported
services to be accessible simultaneously.\footnote{While this issue affects
testing, it rarely affects users, since simply retrying a query is often
enough to fix user issues.  When the servers are simply down or broken,
astroquery is affected, and the resulting errors are sometimes unpredictable;
users are encouraged to report such failures as github issues
(\url{https://github.com/astropy/astroquery/issues}) so that better error
messages can be provided.} We therefore require that each
module provide some tests that do not rely on having an internet connection.
These tests rely on \emph{monkeypatching}\footnote{Monkeypatching is the
  dynamic replacement of attributes at runtime, i.e., changing what
  functions do after they are imported.} to replace the remote
requests. Instead of downloading data, the test suite uses locally available
files to test the query mechanisms and the data parsers.  Monkeypatching in
the context of pytest results in code that is generally more
difficult to understand than typical Python code, but a set of tests
independent of the remote services is necessary.

The local tests are run as part of the continuous integration for the
project with each commit.  The remote tests are run for merges and as part of a
regularly-scheduled cron job.  Running the remote tests less frequently
helps reduce the burden on the remote services.

\subsection{Other utilities}
There are several general-use utilities implemented as part of astroquery, such
as a bulk FITS file downloader and renamer and a download progressbar (these
tools complement similar features in \astropy).  There
is also a schema system implemented to allow user-side parameter validation.
The schema systems are basic syntax-checking tools that verify that the parameters
the user has input are of the right type and format for the target service;
for those services without schemas, the user can hypothetically send queries
that the service will be unable to handle.
The schema tool is only implemented in the ESO and Vizier modules, but it could
be expanded to other modules to reduce the number of doomed-to-fail queries
sent through astroquery.

\section{Development history and status}
\label{sec:development}
Anyone can contribute to astroquery.  The maintainers are committed to helping
developers make new modules that meet the requirements of astroquery.  This
section describes how astroquery has been developed, but we welcome all sorts
of new contributions, including new modules, upgrades to existing modules, and
minor corrections to existing tools from both individuals and institutions.

Astroquery is an \astropy coordinated package \citep{APE15} and is a critical
component of the \astropy Project ecosystem \citep{Astropy-Collaboration2018}.
It is a standalone project and will remain independent of the \astropypkg core
package,\footnote{Many \astropy affiliated packages are developed with the
intent of eventually including them in the core of \astropypkg.  In contrast,
astroquery intends to remain a separate package indefinitely largely because
of its need to rapidly adapt to changes in the remote services; \astropypkg
cannot make such rapid changes because users rely on its stability.} but is
coordinated by the \astropy Project to ensure sustainability and maintenance.

Astroquery has received contributions from 77 people as of August 2018.  While
the primary maintenance burden is shouldered by two people at any given time
(the first two authors), most individual modules have been implemented
independently by interested contributors.

Some contributions have come with direct institutional support.  The ESA Gaia and
ESASky modules were provided and supported by developers working for ESA\@.  The
ADS module is maintained by developers working at ADS. The
MAST and VO Cone Search query tools were added by developers at STScI, with the
latter moved over from \texttt{astropy.vo} (see Section \ref{sec:vo}).

Astroquery also receives contributions from other funded programs. For instance, the
JPLHorizons module has been implemented as part of the \texttt{sbpy}
project\footnote{\url{http://sbpy.org}} with support from NASA. Further Solar
System-related services are planned to be added to astroquery through this
support. Astroquery has also received support from the Google Summer of Code
program, with two students (co-authors Madhura Parikh and Simon Liedtke) from
2013--2014.

Due to its nature as an openly developed package, new directions in
astroquery are primarily driven by contributors and data providers adding or
updating modules to reflect new or changed data sources. The underlying
software architecture has been demonstrably sufficient to meet the needs of
the current generation of data sources (proven by the user base of
astroquery).  While this policy may change in the future, the user-focused
nature of astroquery means that making such architecture changes is
unnecessary until there are specific data sources or use cases to drive
them.

\subsection{Relation to the VO}
\label{sec:vo}

The Virtual Observatory (VO) has some goals similar to astroquery,
though their approach and philosophy is different.  Where VO services provide a
single point of access for all VO-compatible services, astroquery
provides a collection of access points that do not require a specific API from
the hosting service.  The general philosophy in astroquery is to
replicate the web page interface provided by a given service as closely as
possible.  While this approach makes some versions of cross-archive searches
more difficult, it keeps the barrier to entry for new users fairly low and limits
the maintenance burden for upstream developers.

However, there are developments in progress to allow more VO-like queries
within astroquery, such as searching for databases by keywords.  As more
services implement VO-based access, some query modules may adopt VO as a backend,
but these changes should be transparent to users (i.e., the astroquery
interfaces will remain unchanged).  The documentation may guide users on how
to use the more sophisticated VO tools that underly these tools.

Some general VO tools are available in astroquery.  The \texttt{vo\_conesearch}
package, which originally resided in \astropy, is now part of astroquery.  VO
Cone Search has a \texttt{query\_region} interface like the other astroquery
services in addition to the existing interfaces ported over from \astropy.
As of \astropypkg 3.0, \texttt{astropy.vo} no longer exists; therefore,
astroquery is now the primary provider of this VO Cone Search service. From a
typical user's standpoint, switching over from \texttt{astropy.vo} should
result in no difference except for updating their Python \texttt{import}
statements (e.g., \texttt{from astroquery.vo\_conesearch import conesearch}
instead of \texttt{from astropy.vo.client import conesearch}).

\section{Documentation and References}
\label{sec:documentation}
\subsection{Online documentation}
The astroquery modules are documented online and can be accessed  at
\url{https://astroquery.readthedocs.io/}.  We include one detailed example of
how to use astroquery in Appendix \ref{sec:example},
but interested users will find many more on the documentation page and
in the example gallery.\footnote{\url{https://astroquery.readthedocs.io/en/latest/gallery.html}}

\subsection{Other Documents}
Several authors have independently described how to use various astroquery
modules, which is a helpful practice we encourage.
 \begin{itemize}
    \item
        Cosmosim:\footnote{\url{https://www.cosmosim.org/cms/news/cosmosim-package-for-astroquery/}}
        a worked example of downloading data from the cosmosim database,
        including logging in.
    \item \citet{Paletou2014a}: a worked example of querying
        Vizier and SIMBAD to make a surface gravity - effective temperature
        plot for a star survey.
    \item \citet{Guillochon2018a}: the definition of
        the Open Astronomy Catalog API and a description of the astroquery
        module built to use it.
    \item
        MAST:\footnote{\url{https://github.com/spacetelescope/MAST-API-Notebooks/blob/master/AstroqueryIntro/AstroqueryFunctionalityDemo.ipynb}}
        A tutorial on the MAST astroquery interface.
    \item GAIA:\footnote{\url{https://gea.esac.esa.int/archive-help/tutorials/python_cluster/index.html}}
        A tutorial on the GAIA astroquery interface.
\end{itemize}

\section{Summary}
Astroquery is a toolkit for accessing remotely hosted astronomical
data through Python.  It is part of the \astropypkg affiliated package system.
We have described its general layout, its development model, and its role in
developing reproducible workflows.  Astroquery is developed for and by our
community: we welcome any new contributions, and such contributions will continue
to define the future directions of the package.

\acknowledgements

We would like to thank the members of the community that have contributed to
\astroquery, that have opened issues and provided feedback, and have
supported the project in a number of different ways. We are greatful for the
infrastructural support the \astropy community provides.  \astroquery is
supported by and makes use of a number of organizations and services outside
the traditional academic community: GitHub, Travis CI, Appveyor, and Read
the Docs. Our package relies heavily on the following Python dependencies,
we are grateful for their maintainers and contributors: \package{requests}
\package{beautifulsoup}, and \package{keyring}.

We thank Google for financing and organizing the Google Summer of Code
program, that has funded two students (SL, and MP) to work on \astroquery in
2013 and 2014.

The following individuals would like to recognize support for their personal
contributions. BMS is supported by the NSF grant AST-1715122 and
acknowledges support from the DIRAC Institute in the Department of Astronomy
at the University of Washington. The DIRAC Institute is supported through
generous gifts from the Charles and Lisa Simonyi Fund for Arts and Sciences,
and the Washington Research Foundation. MM, MVB, GG contributions are
supported by the NASA PDART grant 80NSSC18K0987.

\software{\astropy \citep{Astropy-Collaboration2018},
  \package{numpy} \citep{numpy},
  \package{requests},
  \package{keyring},
  \package{beautifulsoup4},
  \package{html5lib},
  \package{matplotlib} \citep{matplotlib},
  \package{APLpy} \citep{aplpy},
  \package{pyregions} \citep{pyregions},
  \package{regions} \citep{regions}
}

\bibliographystyle{aasjournal}
\bibliography{bibliography}

\begin{thebibliography}{}
\expandafter\ifx\csname natexlab\endcsname\relax\def\natexlab#1{#1}\fi
\providecommand{\url}[1]{\href{#1}{#1}}
\providecommand{\dodoi}[1]{doi:~\href{http://doi.org/#1}{\nolinkurl{#1}}}
\providecommand{\doeprint}[1]{\href{http://ascl.net/#1}{\nolinkurl{http://ascl.net/#1}}}
\providecommand{\doarXiv}[1]{\href{https://arxiv.org/abs/#1}{\nolinkurl{https://arxiv.org/abs/#1}}}

\bibitem[{{Astropy Collaboration} {et~al.}(2013){Astropy Collaboration},
  {Robitaille}, {Tollerud}, {Greenfield}, {Droettboom}, {Bray}, {Aldcroft},
  {Davis}, {Ginsburg}, {Price-Whelan}, {Kerzendorf}, {Conley}, {Crighton},
  {Barbary}, {Muna}, {Ferguson}, {Grollier}, {Parikh}, {Nair}, {Unther},
  {Deil}, {Woillez}, {Conseil}, {Kramer}, {Turner}, {Singer}, {Fox}, {Weaver},
  {Zabalza}, {Edwards}, {Azalee Bostroem}, {Burke}, {Casey}, {Crawford},
  {Dencheva}, {Ely}, {Jenness}, {Labrie}, {Lim}, {Pierfederici}, {Pontzen},
  {Ptak}, {Refsdal}, {Servillat}, \& {Streicher}}]{Astropy-Collaboration2013}
{Astropy Collaboration}, {Robitaille}, T.~P., {Tollerud}, E.~J., {et~al.} 2013,
  \aap, 558, A33, \dodoi{10.1051/0004-6361/201322068}

\bibitem[{{Astropy Collaboration} {et~al.}(2018){Astropy Collaboration},
  {Price-Whelan}, {Sip{\H o}cz}, {G{\"u}nther}, {Lim}, {Crawford}, {Conseil},
  {Shupe}, {Craig}, {Dencheva}, {Ginsburg}, {VanderPlas}, {Bradley},
  {P{\'e}rez-Su{\'a}rez}, {de Val-Borro}, {Aldcroft}, {Cruz}, {Robitaille},
  {Tollerud}, {Ardelean}, {Babej}, {Bachetti}, {Bakanov}, {Bamford},
  {Barentsen}, {Barmby}, {Baumbach}, {Berry}, {Biscani}, {Boquien}, {Bostroem},
  {Bouma}, {Brammer}, {Bray}, {Breytenbach}, {Buddelmeijer}, {Burke},
  {Calderone}, {Cano Rodr{\'{\i}}guez}, {Cara}, {Cardoso}, {Cheedella},
  {Copin}, {Crichton}, {D{\'A}vella}, {Deil}, {Depagne}, {Dietrich}, {Donath},
  {Droettboom}, {Earl}, {Erben}, {Fabbro}, {Ferreira}, {Finethy}, {Fox},
  {Garrison}, {Gibbons}, {Goldstein}, {Gommers}, {Greco}, {Greenfield},
  {Groener}, {Grollier}, {Hagen}, {Hirst}, {Homeier}, {Horton}, {Hosseinzadeh},
  {Hu}, {Hunkeler}, {Ivezi{\'c}}, {Jain}, {Jenness}, {Kanarek}, {Kendrew},
  {Kern}, {Kerzendorf}, {Khvalko}, {King}, {Kirkby}, {Kulkarni}, {Kumar},
  {Lee}, {Lenz}, {Littlefair}, {Ma}, {Macleod}, {Mastropietro}, {McCully},
  {Montagnac}, {Morris}, {Mueller}, {Mumford}, {Muna}, {Murphy}, {Nelson},
  {Nguyen}, {Ninan}, {N{\"o}the}, {Ogaz}, {Oh}, {Parejko}, {Parley}, {Pascual},
  {Patil}, {Patil}, {Plunkett}, {Prochaska}, {Rastogi}, {Reddy Janga},
  {Sabater}, {Sakurikar}, {Seifert}, {Sherbert}, {Sherwood-Taylor}, {Shih},
  {Sick}, {Silbiger}, {Singanamalla}, {Singer}, {Sladen}, {Sooley},
  {Sornarajah}, {Streicher}, {Teuben}, {Thomas}, {Tremblay}, {Turner},
  {Terr{\'o}n}, {van Kerkwijk}, {de la Vega}, {Watkins}, {Weaver}, {Whitmore},
  {Woillez}, \& {Zabalza}}]{Astropy-Collaboration2018}
{Astropy Collaboration}, {Price-Whelan}, A.~M., {Sip{\H o}cz}, B.~M., {et~al.}
  2018, ArXiv e-prints.
\newblock \doarXiv{1801.02634}

\bibitem[{{Eisner} {et~al.}(2016){Eisner}, {Bally}, {Ginsburg}, \&
  {Sheehan}}]{Eisner2016a}
{Eisner}, J.~A., {Bally}, J.~M., {Ginsburg}, A., \& {Sheehan}, P.~D. 2016,
  \apj, 826, 16, \dodoi{10.3847/0004-637X/826/1/16}

\bibitem[{{Guillochon} \& {Cowperthwaite}(2018)}]{Guillochon2018a}
{Guillochon}, J., \& {Cowperthwaite}, P.~S. 2018, Research Notes of the
  American Astronomical Society, 2, 27, \dodoi{10.3847/2515-5172/aac2c8}

\bibitem[{{Helfand} {et~al.}(2006){Helfand}, {Becker}, {White}, {Fallon}, \&
  {Tuttle}}]{Helfand2006}
{Helfand}, D.~J., {Becker}, R.~H., {White}, R.~L., {Fallon}, A., \& {Tuttle},
  S. 2006, \aj, 131, 2525, \dodoi{10.1086/503253}

\bibitem[{Hunter(2007)}]{matplotlib}
Hunter, J.~D. 2007, Computing In Science \& Engineering, 9, 90,
  \dodoi{10.1109/MCSE.2007.55}

\bibitem[{{Paletou} \& {Zolotukhin}(2014)}]{Paletou2014a}
{Paletou}, F., \& {Zolotukhin}, I. 2014, ArXiv e-prints, arXiv:1408.7026.
\newblock \doarXiv{1408.7026}

\bibitem[{pyregions developers(2018)}]{pyregions}
pyregions developers. 2018, {regions -- ds9 region parser for python},
  \url{https://github.com/astropy/pyregions}

\bibitem[{regions developers(2018)}]{regions}
regions developers. 2018, {regions -- Astropy affiliated package for region
  handling}, \url{https://github.com/astropy/regions}

\bibitem[{{Robitaille} \& {Bressert}(2012)}]{aplpy}
{Robitaille}, T., \& {Bressert}, E. 2012, {APLpy: Astronomical Plotting Library
  in Python}, Astrophysics Source Code Library.
\newblock \doeprint{1208.017}

\bibitem[{Tollerud(2018)}]{APE15}
Tollerud, E. 2018, {Astropy Proposal for Enhancement 15: An Updated Model for
  the Affiliated Package Ecosystem (APE 15)}, \dodoi{10.5281/zenodo.1246834}.
\newblock \url{https://doi.org/10.5281/zenodo.1246834}

\bibitem[{{Van der Walt} {et~al.}(2011){Van der Walt}, Colbert, \&
  Varoquaux}]{numpy}
{Van der Walt}, S., Colbert, S.~C., \& Varoquaux, G. 2011, {Computing in
  Science \& Engineering}, 13, 22,
  \dodoi{http://dx.doi.org/10.1109/MCSE.2011.37}

\bibitem[{{Vogt}(2018)}]{Vogt2018a}
{Vogt}, F.~P.~A. 2018.
\newblock \doarXiv{1807.02114v1}

\end{thebibliography}

\appendix
\section{Example}
\label{sec:example}
In this appendix, we show an example of astroquery in action, highlighting the
ability to use multiple modules and interact with \astropypkg's table, coordinate,
and unit tools.  This example approximately reproduces Figure 1 of
\citet{Eisner2016a}, but with a different background.  It can also be found on
astroquery's gallery page
(\url{http://astroquery.readthedocs.io/en/latest/gallery.html}).
Another illustration of how to use astroquery tools in a finder chart making
tool is \texttt{fcmaker}, which produces charts for ESO observations using
astroquery \citep{Vogt2018a}.

\newpage
\begin{lstlisting}[frame=single]
# Create a finder chart and overlay two catalogs using the Vizier and SkyView
# tools
from astropy import units as u
from astropy.coordinates import SkyCoord
from astropy.wcs import WCS
from astroquery.skyview import SkyView
from astroquery.vizier import Vizier
import matplotlib.pyplot as plt

center = SkyCoord.from_name("Orion KL")

# Grab an image from SkyView of the Orion KL nebula region
imglist = SkyView.get_images(position=center, survey="2MASS-J")

# The returned value is a list of images, but there is only one
img = imglist[0]

# "img" is now a fits.HDUList object; the 0th entry is the image
mywcs = WCS(img[0].header)

fig = plt.figure(1)
fig.clf()  # Just in case one was open before
# Use astropy's wcsaxes tool to create an RA/Dec image
ax = fig.add_axes([0.15, 0.1, 0.8, 0.8], projection=mywcs)
ax.set_xlabel("RA")
ax.set_ylabel("Dec")

ax.imshow(img[0].data, cmap="gray_r", interpolation="none", origin="lower",
          norm=plt.matplotlib.colors.LogNorm())

# Retrieve a specific table from Vizier to overplot
tablelist = Vizier.query_region(
    center, radius=5*u.arcmin, catalog="J/ApJ/826/16/table1")
# Again, the result is a list of tables, so we"ll get the first one
result = tablelist[0]

# Convert the ra/dec entries in the table to astropy coordinates
tbl_crds = SkyCoord(result["RAJ2000"], result["DEJ2000"],
                    unit=(u.hour, u.deg), frame="fk5")

# We want this table too:
tablelist2 = Vizier(row_limit=10000).query_region(
    center, radius=5*u.arcmin, catalog="J/ApJ/540/236")
result2 = tablelist2[0]
tbl_crds2 = SkyCoord(result2["RAJ2000"], result2["DEJ2000"],
                     unit=(u.hour, u.deg), frame="fk5")

# Overplot the data in the image
ax.plot(tbl_crds.ra, tbl_crds.dec, "*", transform=ax.get_transform("fk5"),
        mec="b", mfc="none")
ax.plot(tbl_crds2.ra, tbl_crds2.dec, "o", transform=ax.get_transform("fk5"),
        mec="r", mfc="none")
# Zoom in on the relevant region
ax.axis([100,200,100,200])

plt.show()
\end{lstlisting}

\begin{figure*}[!htp]
\includegraphics[scale=1,width=7in]{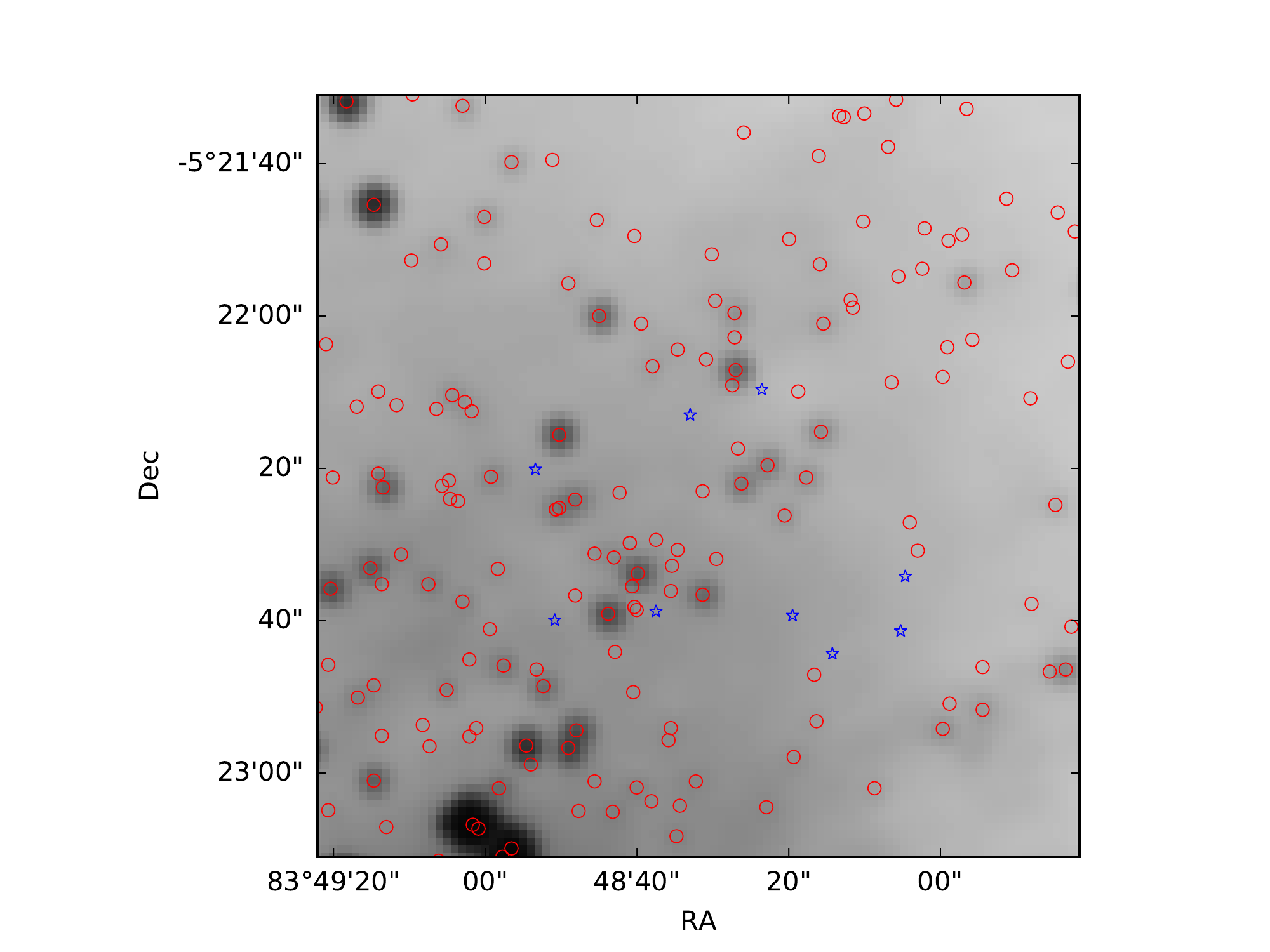}
\caption{An example figure made using astroquery.  The \texttt{skyview} package
was used to download a 2MASS J-band image.  The \texttt{vizier} was used to
download two star catalogs from different publications and overplot them; the
blue stars show sources from the older, less complete catalog and the red
circles show sources from a more recent, more complete catalog.
}
\label{fig:example1}
\end{figure*}

\end{document}